\documentclass{kluwer}
\usepackage{graphicx}
\begin{document}
\begin{article}
\begin{opening}
\title{Molecular absorption in Cen~A on VLBI scales}

\author{Huib Jan \surname{van Langevelde}\email{langevelde@jive.nl}}
\institute{Joint Institute for VLBI in Europe, 
Postbus 2, 7990 AA, Dwingeloo, the Netherlands}
\author{Ylva \surname{Pihlstr\"om}}
\institute{National Radio Astronomical Observatory, 
PO Box 0, Socorro NM~87801, U.S.A.}
\author{Anthony \surname{Beasley}}
\institute{CARMA/California Institute of Technology
Owens Valley Radio Observatory, Big Pine, CA~91125, U.S.A.}

%\date: rather not

\runningtitle{Molecular absorption in Cen~A}
\runningauthor{van Langevelde et al.}

%\begin{ao}
%Joint Institute for VLBI in Europe\\
%Postbus 2\\
%7990 AA Dwingeloo\\
%The Netherlands
%\end{ao}

\begin{abstract}
  Centaurus A, the nearest AGN shows molecular absorption in the
  millimeter and radio regime. By observing the absorption with VLBI,
  we try to constrain the distribution of the gas, in particular
  whether it resides in the circumnuclear region. Analysis of VLBA
  observations in four OH and two H$_2$CO transitions is presented here, as
  well as molecular excitation models parameterized with distance from
  the AGN. We conclude that the gas is most likely associated with the
  tilted molecular ring structure observed before in molecular emission
  and IR continuum. The formaldehyde absorption shows small scale
  absorption which requires a different distribution than the hydroxyl.
\end{abstract}


\end{opening}

\section{Introduction}
Centaurus A, the radio source in the nearest AGN, displays remarkable
absorption in both molecular and atomic gas. While in similar objects the
absorption is often made up of a few broad components, in Cen~A it
can be resolved into many narrow components that together cover 80 km/s.
Moreover, these features can be traced in very many chemical
components, because Cen~A has a bright nucleus from radio to
sub-millimeter wavelengths.

A matter of debate has been whether these tracers are associated with
the obscuring dust band in the outer regions of NGC~5128, or whether
some of these originate in the circumnuclear region. The first case
would allow us to study the physics and chemistry of the interstellar
medium in a recent merger galaxy, undergoing a starburst. The second
case would possibly be even more exciting, as these lines could then be
used as tracers of the physical and chemical conditions in the vicinity
of a black hole, where conditions are affected by intense radiation
and the ultimate fate of the gas is to feed the monster.

There are at least two ways in which we can hope to determine the
location of the absorbing gas on the line of sight. The first is
through geometrical arguments; for example, if it can be shown that
only a small fraction of the background source is obscured and a torus
shape is assumed, it may be argued that the obscuring ring is close to
the nucleus.

The second method is to investigate whether the molecular excitation
requires or excludes the influence of the nuclear emission. A related
argument may be the chemical composition of the material, which could
show the signature particular to X-ray processing if the gas is of
nuclear origin.

In this paper we combine VLBA observations of OH and H$_2$CO with excitation
modeling to argue that most of the molecular material seen against Cen~A
is at considerable distance from the nucleus (between 200 pc to 2 kpc).

\section{VLBA results}

\begin{figure} % figuur 1
\includegraphics[angle=-90,width=\linewidth]{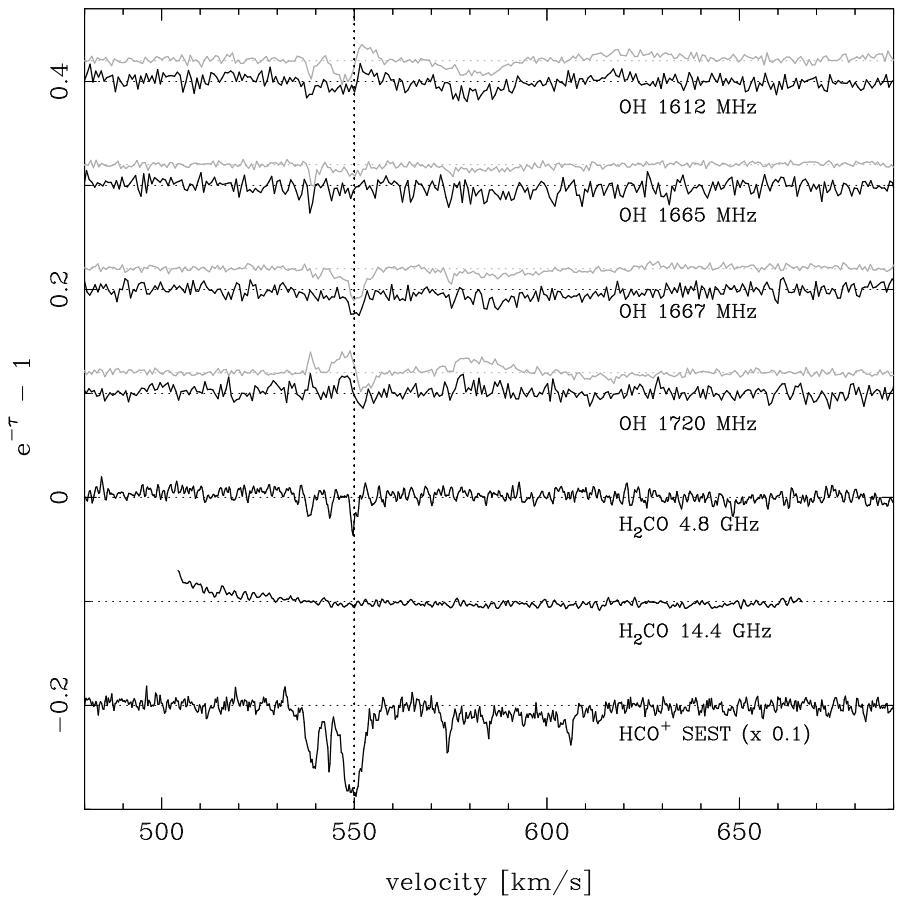}
%\vspace{6pc}
\caption[]{Spectra for OH and H$_2$CO from VLBA data, in a sub-arcsecond
beam. For comparison OH spectra obtained with the ATCA are shown as well
in grey, slightly offset from the VLBA data \cite{vanlangevelde}.
The bottom spectrum is HCO$^+$ from the SEST \cite{israelabs}.}
\label{spec}
\end{figure}

The VLBA OH 18cm data was taken on July 8 1995.  A single track was
observed with all 10 telescopes. The frequency setup covered four
base-bands in two polarizations, covering a bandwidth of 2 MHz
($\approx 360$km/s). In order to obtain sufficient spectral resolution
($\approx 0.8$km/s), two correlator passes were done. The formaldehyde
lines were initially observed on 12 December 1996, covering the 6cm and
2cm transitions on consecutive days. Analysis of these observations
showed that the 6cm data had some problems and those were re-observed
on November 20 1997.

\begin{figure}
  \centering
  \includegraphics[angle=0,width=0.49\textwidth]{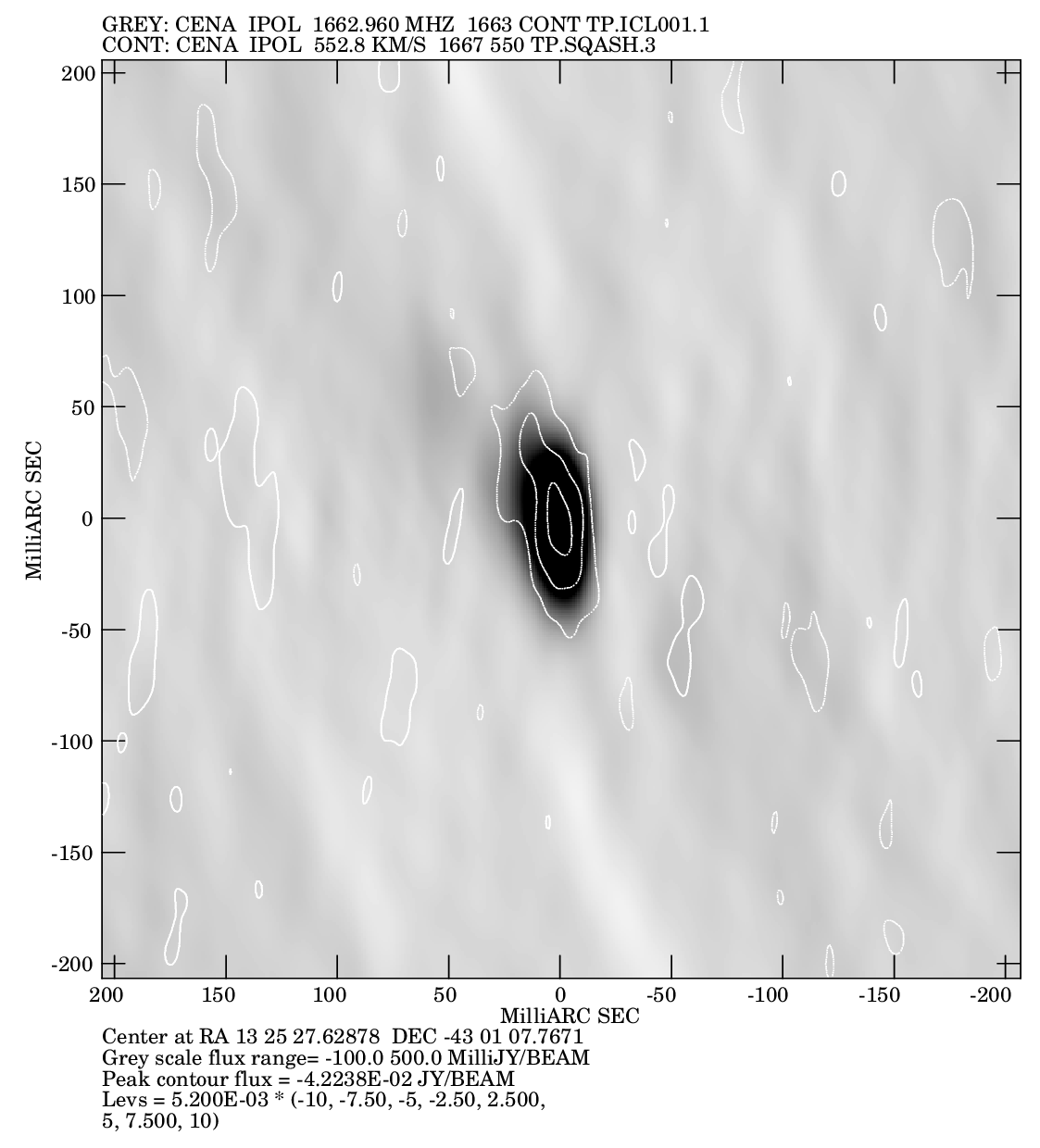}
  \includegraphics[angle=0,width=0.49\textwidth]{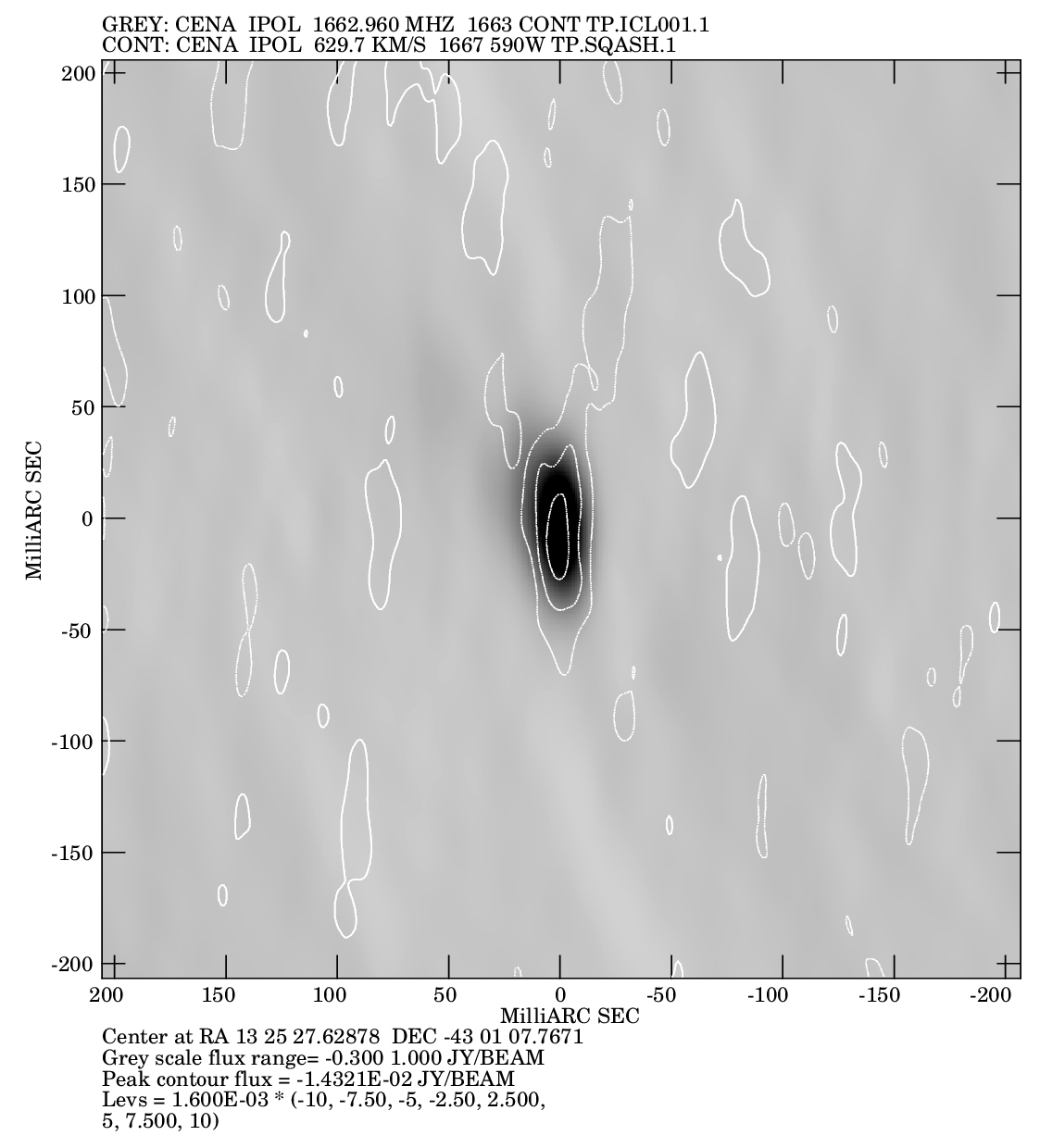}
\caption{Contours show the OH 1667 MHz absorption feature at 550 km/s (left),
  overlayed on the continuum in grey scale, as well as the OH 1667
  feature in the 590 km/s shallow component. The absorption seems to
  cover a large fraction of the source, at least at 550 km/s.}
\label{ohmap}
\end{figure}

It turned out to be quite complicated to image the data properly; the
low elevations at which Cen~A is observed from the VLBA sites imply
that the mutual visibility is limited. The array of antennas that
has data on Cen~A and can be calibrated properly changes many times
during the stretch of the observations. This makes self-calibration quite
difficult, especially for the 18cm data which is dominated by flux on short
baselines.

As a result, the best spectra are obtained directly in the uv-domain
and show the most signal to noise when limited to short baselines.
These spectra are shown in Figure~\ref{spec}.  They are compared with a
SEST HCO$^+$ spectrum \cite{israelabs} and previous ATCA results
\cite{vanlangevelde}. The resulting VLBA OH spectra match the previous
ATCA results, if we take the higher signal to noise in the large beam
into account. This indicates that the absorption is smooth between the
ATCA and VLBA scales. Earlier results on H$_2$CO were obtained by
\citeauthor{seaquistbell} \shortcite{seaquistbell}, who detected both
formaldehyde lines with the VLA. Again the VLBA formaldehyde results
seem consistent with their work, given the lower signal to noise in the
high resolution data.

The next step is to image the hydroxyl absorption against the VLBI
structure of Centaurus A. This is shown in Figure~\ref{ohmap}.  At a
distance of 3.4 Mpc 50 mas corresponds to 0.82pc \cite{israelrev}.  It is
clear that the systemic feature covers the entire source.
Interestingly, at the red-shifted velocity it appears at first that the
absorption only covers the brightest part of the source, but is not
detected against the extended jet. However, careful analysis of the
signal to noise in the source shows that the lack of absorption is not
significant; the background is simply not bright enough to detect a
homogeneous absorption screen at the jet position. We conclude that the OH
result is consistent with both absorption components covering the
entire VLBI source. But it is equally consistent with the systemic
component covering the entire source and a red-shifted component in
front of the brightest 18cm component.

\begin{figure}
  \centering
  \includegraphics[angle=0,width=0.6\textwidth]{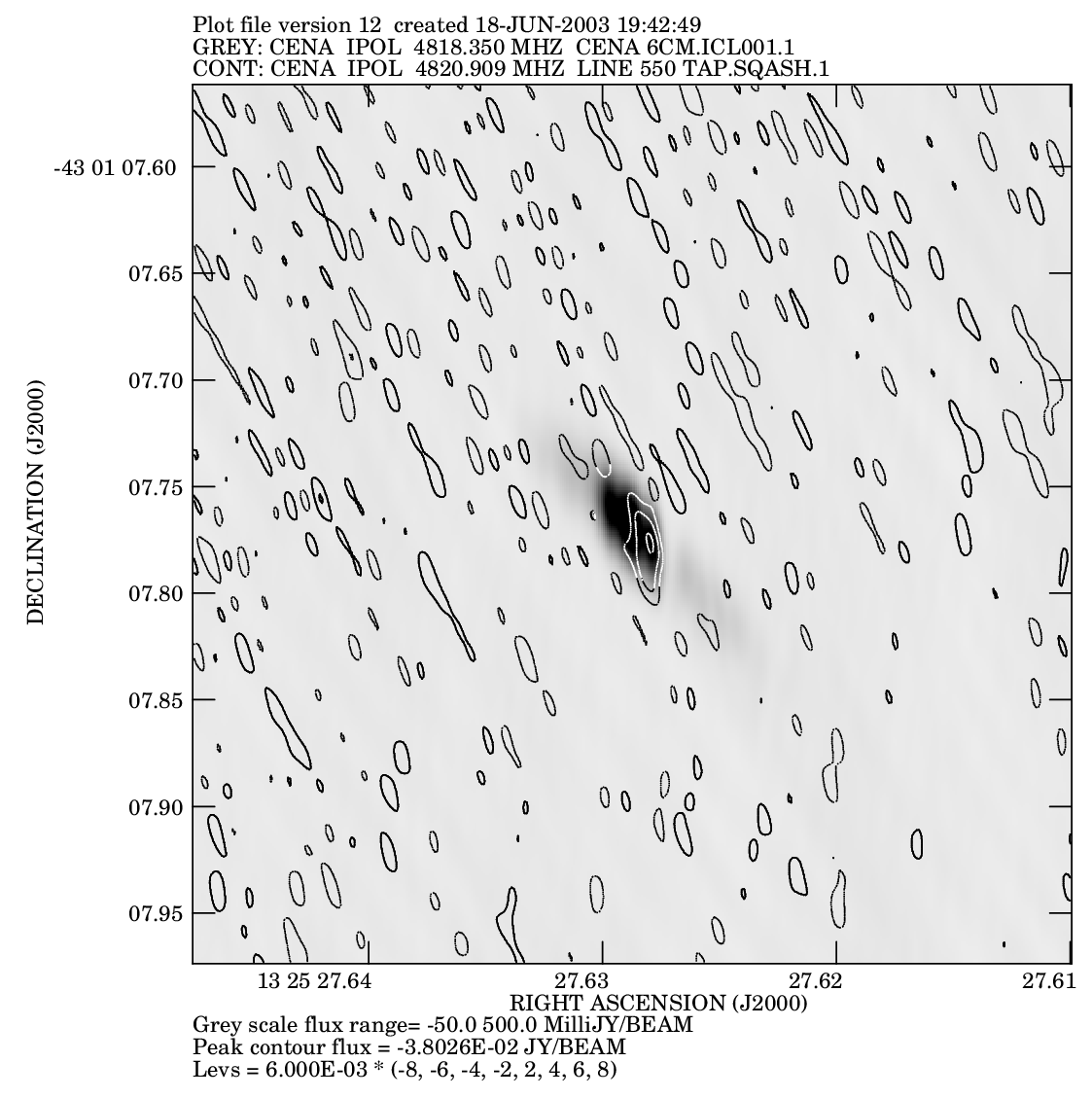}
\caption{A map of the 4.8 GHz continuum of Cen~A overlayed
with the H$_2$CO absorption in the 550 km/s feature. Clearly the
feature does not cover the entire structure equally.}
\label{formmap}
\end{figure}

In Figure~\ref{formmap} the results from mapping the formaldehyde are
shown. It should be kept in mind that at the 4.8 (and 15 GHz) we may be
looking at different components in continuum emission.  Following the
analysis by \citeauthor{tingaymurphy} \shortcite{tingaymurphy} we
interpret the brightest spot in the 1.6 and 4.8 GHz image as a knot in
the jet. In the 18cm image the core is absorbed, at 6cm the core and
jet components are approximately equally bright. We detect formaldehyde
absorption at the systemic velocity against the core component.
Surprisingly there is no absorption detected against the knot in the
jet. As this component is equally bright as the core, {\sl this}
structure is significant. We conclude that the systemic formaldehyde
absorption arises from a different place than the OH component, even
though the velocity is identical.

\section{Modeling}

We have modeled the excitation of the OH and H$_2$CO following the
analysis in \citeauthor{vanlangevelde} \shortcite{vanlangevelde}.  The
gas in front of Cen~A is treated with a simple escape probability code.
In this analysis we have put emphasis on the excitation of the
molecular gas by external radiation, expected to be an important factor
in a galaxy that shows both signatures of starburst and an active
nucleus. The excitation of both species is governed by the usual
parameters of kinetic temperature, density of the ambient H$_2$ and
OH/H$_2$CO column density, which transfers into a length scale and an
abundance. In addition the excitation of OH by infrared radiation is
taken into account. Following the previous analysis we consider two
components: far-infrared radiation generated at the location of the OH
gas, as well as intense mid-infrared radiation from the inner
circumnuclear disk ($<$ 400pc).  For the excitation of formaldehyde
radio emission is important and therefore a flat spectrum nuclear
source is included too.

The total model has far too many parameters to explore all
dependencies.  Rather, we have fixed all our best guesses based on the
available observations of Cen~A and varied only one physical
constraint, namely the distance of the absorbing gas from the nucleus
of Cen~A.  Keeping the density and temperature of the gas constant, as
well as the locally generated far infrared radiation, the molecular gas
is subject to less mid-infrared and radio excitation as the solid angle
that the nuclear source subtends decreases with increasing distance
from the center (and for radii within the molecular ring the
mid-infrared emission is assumed to be local). The relative strength of
the 1.6~GHz OH components and the dominance of the 4.8~GHz formaldehyde
over the 15~GHz line are taken to be the main quantities to be
reproduced.

Within the context of this simple model, analysis shows that the gas
cannot be much closer than 200pc. Otherwise we predict maser emission
in the 6cm formaldehyde line, as the radio emission starts to dominate
the excitation.  A similar effect appears in the OH models, where the
infrared emission will dominate the excitation and main line masers
will occur. It is more difficult to rule out very large distances from
the nucleus, anything up to 2 kpc is possible, showing that this
excitation is not exceptional compared to normal Galactic conditions.
At very large distances the main lines are expected to be much stronger
than the satellite lines. This does not appear in the main systemic
component, but it seems this effect can be seen at some of the weaker
systemic components that are slightly blue-shifted.

\section{Discussion}

From the excitation analysis it follows that the OH and formaldehyde
gas seems to be constrained to radii of 200pc - 2kpc. This seems to be
a robust results based on interpretation of global properties of the
spectra, although it should be realized that this follows from a rather
crude model. Shielding of the molecular gas, e.g.\ in a very thin disk,
could possibly allow a closer radius. On the other hand, a strong local
interstellar radiation field, possible in starburst conditions, could
support the specific excitation at larger radii.

On the relevant scales many authors have studied the structure and
physics of the gas and dust in NGC~5128. Much of the discussion of on
these components can be found in \citeauthor{israelrev}
\shortcite{israelrev}, but for our interpretation a model by
\citeauthor{eckartmod} \shortcite{eckartmod} is most relevant. This
model is based on CO, sub-mm and infrared imaging (e.g.\ 
\citeauthor{leeuw} \citeyear{leeuw}; \citeauthor{wild} \citeyear{wild})
which shows a warped structure on the scale of 0.5 -- 1.5 kpc.
Modeling this structure as a set of concentric disks, each with a
slightly different axis orientation {\sl and} assuming gas at
considerable height above these is co-rotating, the millimeter
absorption structure (like the HCO$^+$ spectrum in Fig 1) can be
reproduced remarkably well.

There seems to be no problem to interpret the OH results in this
context.  According to this model neutral molecular gas components
close to the systemic velocity arise at very different distances from
the nucleus. At three different radii the tilted rings intersect the
line of sight to the nucleus. This could thus be consistent with the
different excitation of the systemic components.  The red-shifted
components arise at higher latitude above the tilted disk in this
model.  Like HCO$^+$ there must thus be considerable OH in the medium
above the mid-plane. As this gas may be less embedded in the high
density mid-plane of the disks one may expect more pronounced
excitation effects. Indeed the satellite lines dominate in this
component.  Compared to the HCO$^+$, the OH seems to be more diffuse
and shows fewer distinct components. The VLBI source extends for less
than 1 pc. With a scale height of 300pc it is therefore expected that
OH absorption is homogeneous over the source, as observed.

The interpretation of the formaldehyde is less straightforward. From
the spectrum it is noticeable that no formaldehyde is seen in the
red-shifted component. Also the systemic components are much narrower.
Moreover, the absorption occurs only against the core of the radio
source, a line of sight not even sampled in OH.

If we interpret the H$_2$CO absorption structure across the source as a
thin disk around the central engine, we derive a distance from the
nucleus which is disturbingly small. At a distance of 3.4 Mpc the size
of the absorption amounts to 0.4pc. Even if this arises from a very
thin disk, with a height/radius ratio of 100, this disk would be
located within the 50 pc from the nucleus. Given the excitation modeling, as
well as the narrow width of the absorption lines this seems unlikely.

An interpretation in which the formaldehyde absorption is located at
larger distance runs into the uncomfortable situation that it requires
a chance alignment of the absorbing cloud in front of the nucleus.
Within the context of the model of \citeauthor{eckartmod}
\shortcite{eckartmod}, it is possible to explain the observations if
the bulk of H$_2$CO is restricted to small clouds ($<1$pc) in the
mid-plane of the tilted rings. This fits in with the absence of
red-shifted components in this tracer, the narrow line width and the
small-scale structure in the absorption. However, on the relevant scale
it is not clear why this absorption is seen against the core only. It
is expected that the other velocity components could equally well
appear in front of the jet. Unfortunately we lack the signal-to-noise
to test the appearance of the weaker components.

We conclude that the VLBI observations, as well as the excitation
modeling is consistent with the absorption occurring on 200 -- 2000 pc
from the center of Centaurus A. The formaldehyde seems to have a
distinctly different distribution, possibly restricted to small, high
density clouds in the circumnuclear disk.

\end{article}
\end{document}